\documentclass[prl,twocolumn]{revtex4}
\usepackage{dcolumn}
\usepackage{amsmath}
\usepackage{graphicx}

\usepackage[
colorlinks=true, 
pdfstartview=FitV, 
linkcolor=blue, 
citecolor=magenta, 
urlcolor=blue, 
bookmarks=true,
bookmarksnumbered=true,
pdftitle={},
pdfauthor={}
]{hyperref}

\newcommand{\pder}[2]{\frac{\partial #1}{\partial  #2}}
\newcommand{\der}[2]{\frac{d #1}{d  #2}}
\newcommand{\para}[1]{{\em #1}\/.---}
\newcommand{\Fc}{F_{\rm c}}

\begin{document}
\title{Derivation of the Invariant Free-Energy Landscape Based on Langevin Dynamics }

\author{Takenobu Nakamura}
\email{takenobu.nakamura@aist.go.jp}
\affiliation{
National Institute of Advanced Industrial Science and Technology (AIST),
1-1-1, Umezono, Tsukuba, Ibaraki 305-8568, Japan}
\affiliation{
JST, PRESTO,
4-1-8 Honcho, Kawaguchi, Saitama 332-0012, Japan}

\begin{abstract}
Conventionally defined free-energy landscape (FEL) exhibits unphysical dependence on the choice of reaction coordinates and hence lacks universal predictive ability.  
We here show that three physically plausible requirements uniquely determine the FEL formula for a given reaction coordinate.
Our FEL is expressed solely in terms of quantities obtained through time-series data analysis, namely, the probability distribution and the diffusion matrix.
It is free from any unphysical coordinate dependence and coincides with the conventional FEL in special cases.
The uniqueness and robustness of the formula strongly suggest that our FEL has universal predictive power.
\end{abstract}

\maketitle
\para{Introduction}
The free-energy landscape (FEL) is regarded as a powerful tool for describing the stability, reaction spontaneity, reaction path, transition state, and energetics of small systems in which thermal fluctuations play a dominant role \cite{sekimoto2010stochastic}. 
The FEL can provide useful insights into biomolecular systems, chemical reactions, and the dynamics of cluster glasses \cite{wales2003energy,chipot2007free} and enables one to describe the states and dynamics of a system near equilibrium solely in terms of a small number of coordinates \cite{zuckerman2010statistical}.
These coordinates may represent an internal or local structure of the molecule, such as a bond length, bond orientational order, bond angle, or dihedral angle, for example. 
Depending on the context, these coordinates may be referred to as ``reaction coordinates,'' ``collective variables,'' ``slow variables,'' or ``coarse-grained variables.'' In this paper, we refer to them simply as reaction coordinates.

Although there have been numerous applications of the FEL, it is known that there exists a fundamental problem: 
the conventionally defined FEL and its physical predictions depend explicitly on the choice of reaction coordinates, and there are no criteria for selecting a unique definition.
It was thus claimed that ``it is not meaningful to speak of {\em the}\/ free-energy landscape'' \cite{frenkel2013simulations}. 
In this Letter, we derive a formula for FEL based on physical assumptions, instead of proposing a definition.
The formula makes it possible to speak of {\em the}\/ FEL.

\para{Motivation and the Main Result}
The FEL is an effective Hamiltonian describing a small subsystem within a larger system under equilibrium conditions.
Suppose, for example, that one is interested in the behavior of a
protein in a solvent.
In this case, the whole system consists of the protein and the
solvent and the subsystem consists of the protein alone.
Let $\Gamma$ be a collection of positions and momentums of the protein and the solvent in the microscopic description.
The equilibrium property is described by the Hamiltonian $H(\Gamma)$.
The reaction coordinates $z$, which are function $z(\Gamma)$ of $\Gamma$, express the internal degree of freedom of the protein such as a collection of dihedral angles or relevant principal components.
Then the FEL should be a "Hamiltonian" expressed as a function of $z$ that characterizes the property of the protein or the subsystem in general.


Our main result is the derivation of the following formula for the FEL $F(z)$ based on the physical assumptions that the FEL must satisfy:
\begin{align}
F(z)=-T\ln P(z)\sqrt{\det {\bf D}(z)}+{\rm const.}
\label{FEL}
\end{align}
Here, $P(z)$ denotes the equilibrium probability density, ${\bf D}(z)$ denotes the diffusion matrix, and $\det {\bf D}(z)$ is its determinant.  
We note that both $P(z)$ and ${\bf D}(z)$ can be fully determined from experimental (or numerical) data. 
The introduction of the diffusion matrix is one of the essential key points in our construction.
We stress that our formula Eq. \eqref{FEL} is not only logically derived but also serves as a basis for practical applications. 
It provides a reliable foundation of the FEL and also practical methods for obtaining the FEL in a variety of systems mentioned above by means of time-series analysis of experimental or simulated data.

\para{Definition and Problem in Conventional FEL}
The conventional FEL is defined by a straightforward coarse-graining procedure as follows.
Assume the whole system $\Gamma$ is in equilibrium at temperature $T$.
Then, the probability distribution of $\Gamma$ is expressed as a canonical distribution ${\cal P}(\Gamma)\propto e^{-H(\Gamma)/T}$.
The conventional definition requires the same relation for the FEL $\Fc(z)$ and probability density distribution $P(z)\propto e^{-\Fc(z)/T}$, i.e,
\begin{align}
\Fc(z)\equiv-T\ln P(z)+{\rm const.},
\label{def1}
\end{align}
where the probability distribution of $z$ is obtained as the marginal distribution i.e., the coarse-grained distribution of $P(\Gamma)$, i.e.,
\begin{align}
P(z)=\int d\Gamma\, \delta(z-z(\Gamma)){\cal P}(\Gamma).
\label{marginal}
\end{align}
Thus, the definition of $\Fc(z)$ comes from the consistency of the two descriptions of equilibrium state by $\Gamma$ and $z$.

As mentioned, $P(z)$ is evaluated through time-series analysis rather than using the definition Eq. \eqref{marginal} itself.
Typically, time-series data is obtained by an MD simulation as $z_t=z(\Gamma_t)$ or obtained by the sequential experimental data.
When the dynamics of the reaction coordinates can be regarded as a stationary ergodic process, we have an alternative expression,
\begin{align}
P(z)=\langle \delta(z-z_t)\rangle, 
\end{align}
for the probability density in Eq. \eqref{marginal}, where $\langle \cdot \rangle$ denotes the time average in the stationary process. 
This is indeed the basis of the histogram method, which is frequently used to compute $\Fc(z)$ from molecular dynamics simulations \cite{chipot2007free} and experimental data analyses \cite{mccann1999thermally}.

In light of our conclusion that \eqref{FEL} is the unique formula for FEL that satisfies the physically necessary invariance assumption, it is clear that the conventional FEL \eqref{def1} lacks the invariance. 
For completeness, let us explicitly see the problems with \eqref{def1}. 
To this end, take another set of reaction coordinates $z'$ that is related to $z$ by $z'=z'(z)$, where $z'(\cdot)$ is a smooth one-to-one function.
Clearly, the coordinates $z$ and $z'$ are equivalent in the sense that they have exactly the same physical information about the system.
Now, note that the probability distribution function for $z'$, $P'(z')=\int d\Gamma\, \delta[z'-z'(z(\Gamma))]{\cal P}(\Gamma)$, is related to $P(z)$ by
\begin{align}
P'(z')=P(z(z'))\Bigl|\pder{z}{z'}\Bigr|.
\label{pdfjacobian}
\end{align}
Therefore the corresponding FEL is given by
\begin{align}
\Fc'(z')&=-T\ln P'(z')+{\cal F}_T\notag\\
&= \Fc(z(z'))-T\ln\Bigl|\pder{z}{z'}\Bigr|,
\label{ambiguity}
\end{align}
which shows that $\Fc'(z')$ and $\Fc(z(z'))$ may differ in general.
We thus conclude that the landscapes described by $z$ and $z'$ are in general different.
This may not be physical because the actual landscape (if it exists) should not depend on the choice of coordinates.
It should be stressed that the different FELs indeed lead to different predictions of energy balance about the behavior of the reaction coordinates.
Note that the Jacobian contribution in Eq. \eqref{ambiguity} does not come from intrinsic properties of the system; rather, it originates from our choice of transformation. 
This shows that there is a serious problem with the reliability of the free-energy barriers 
or the reaction paths obtained from numerical simulations \cite{frenkel2013simulations,jungblut2016pathways} or experimental data \cite{mccann1999thermally}. 
In this way, the coordinate transformation reveals the inadequacy of the conventional FEL.

Note that Eq. \eqref{ambiguity} implies that $\Fc'(z')=\Fc(z(z'))$ when $T=0$.
This is consistent with the fact that the energy landscape and intrinsic reaction coordinate 
are well-defined quantities, free from the Jacobian problem \cite{fukui1970formulation}.
Note also that one has $\Fc'(z')\simeq\Fc(z(z'))$ in a macroscopic system because $\Fc$ is proportional to the volume of the system while the Jacobian $|\partial z/\partial z'|$ is of order 1.
We see that the above problem of the coordinate dependence of the landscape is intrinsic to small systems at nonzero temperatures.

To overcome the problem with the conventional FEL defined as Eq. \eqref{def1}, we do not try to propose an arbitrary "better" definition but shall derive a unique formula for the FEL based on three physically plausible requirements.  
Our derivation is based on a dynamical requirement that the FEL describes the reaction, physical and geometric requirement about the "force" that we shall define below,
and the consistency with the canonical distribution. 
The justification for these requirements is provided below.

\para{Requirement 1: Dynamics}
First, let us reconsider the purpose of the FEL.
The main objective is to evaluate the reaction rate and extract the reaction path described by the physical state specified by $z$.
The reaction rate is essentially a dynamical quantity; thus, dynamics must be specified explicitly.
The dynamics must be stochastic because the degrees of freedom are reduced.
Therefore, we need to start with a stochastic process.
This stochastic process must not include memory effects and thus must be Markovian to describe the reaction rate only in terms of the FEL.
It is worth mentioning that some non-Markovian models can be embedded into the high-dimensional Markovian model \cite{kawai2015environmental}, thus this assumption would be applicable to some non-Markovian models.
The Markov process must also be continuous because the reaction path is represented as a trajectory.
In this way, the over-damped Langevin equation needs to be specified as a starting point for discussing the FEL.
More explicitly, if the FEL is to be interpreted according to the Kramers rate formula, the description by the over-damped Langevin equation, which is the starting point of the formula, is essential.

This requirement does not mean that the microscopic dynamics must obey the Langevin equation.
The microscopic dynamics may be Hamiltonian dynamics represented by $\Gamma_t$ with the Hamiltonian $H(\Gamma)$.
Instead, we require that, irrespective of the microscopic dynamics, there is a separation of the time scales that justifies the description by the Langevin equation \cite{itami2017universal}.

First, let us observe how the Langevin equation changes according to a coordinate transformation.
For simplicity, we assume a Langevin equation with additive noise and $n=1$, i.e., $z$ has a single component. 
Then, the time-evolution equation of $z$ is given as
\begin{align}
\der{z_t}{t}= A(z_t)+S_0 R_t.
\label{additive}
\end{align}
Here, $A(z)$ is a function of $z$ and $S_0$ is a constant.
$R_t$ is a zero-mean Gaussian process satisfying
$\langle R_tR_{t'}\rangle =2\delta(t-t')$. 
Here, and in what follows, stochastic differential equations are interpreted in the sense of Stratonovich \cite{risken1989fpe}.

Then, let us perform a coordinate transformation of variable $z$ into $z'=z'(z)$.
By using the chain rule, we obtain the time-evolution equation of $\bar z$ as
\begin{align}
\der{z'_t}{t} = A'(z'_t)+S'(z'_t) R_t,
\label{multiplicative}
\end{align}
where $A'(z')$ and $S'(z')$ are functions of $z'$.
As we have seen here, the Langevin equation with additive noise, such as Eq. \eqref{additive}, is not invariant with respect to coordinate transformations. 
To be invariant to coordinate transformations, we must start with the Langevin equation in a form that includes multiplicative noise such as Eq. \eqref{multiplicative}.

Let us move on to the general case
with $n$ reaction coordinates. 
For simplicity, we only treat the even variables with respect to time reversal.
To begin with, we introduce some fundamental equations and quantities.
The over-damped Langevin
equation with multiplicative noise corresponding to
Eq. \eqref{multiplicative} takes the form
\begin{align}
\der{z^i_t}{t}=A^i(z_t)+\sum_{\mu}S^{i\mu}(z_t)R^\mu_t,
\label{normal form}
\end{align}
where $R_t^i$ is the zero-mean Gaussian process satisfying $\langle R_t^\mu R_{t'}^\nu\rangle=2\delta^{\mu\nu}\delta(t-t')$.
Here, $A^i(z)$ and $S^{i\mu}(z)$ are not directly determined from a time-series analysis.
Rather, the drift vector ${\boldsymbol v}(z)=(v^i(z))_{i=1,2,...,n}$ and the diffusion matrix 
${\bf D}(z)=(D^{ij}(z))_{i,j=1,2,...,n}$ can be 
read off from a stationary process of reaction coordinates $z_t$ as\begin{align}
v^i(z)
&=\lim_{\tau \to 0}\frac{\langle z^i_{t+\tau}-z^i\rangle_{z_t=z}}{\tau},
\label{def:drift}\\
D^{ij}(z)&=\lim_{\tau \to 0}\frac{\langle (z^i_{t+\tau}-z^i)(z^j_{t+\tau}-z^j)\rangle_{z_t=z}}{2\tau}.
\label{def:diffusion}
\end{align}
Here, $\langle \cdot \rangle_{z_t=z}$ denotes the conditional average given $z_t=z$ (see \cite{risken1989fpe}).
The above quantities such as $S^{ij}(z)$, $D^{ij}(z)$, $A^i(z)$, and $v^i(z)$ are related with each other as follows: 
\begin{align}
v^i(z)&=A^i(z)+
\sum_{\mu,k} S^{k\mu}(z)\pder{S^{i\mu}(z)}{z^k},\\
D^{ij}(z)&=\sum_{\mu} S^{i\mu}(z)S^{j\mu}(z).
\label{StoD}
\end{align}
By definition, ${\bf D}(z)$ is 
a positive semi-definite symmetric matrix 
and a second-rank contravariant tensor \cite{graham1977covariant,berezhkovskii2011time}.
We further assume that it is positive definite so that its inverse exists.

These quantities appear in the Fokker--Planck equation obtained by rewriting the Langevin equation as the time evolution of the probability density $P(z,t)$ at time $t$, expressed as
\begin{align}
\pder{P(z,t)}{t}=-\pder{}{z^i}\left[\left(v^i(z)-\pder{}{z^j}D^{ij}(z)\right)P(z,t)\right].
\label{FokkerPlanck}
\end{align}
Here, and in what follows, we use Einstein's convention for the sum of indices.
This equation represents the transition probability.
Because the stochastic dynamics are completely characterized in terms of the transition probability, only the measurable quantities $v^i(z)$ and $D^{ij}(z)$ determine the physical quantities \cite{hinczewski2010diffusivity}.
Therefore, the FEL must also be expressed through them.

\para{Requirement 2: Covariance}
Next, let us reconsider how the FEL should be expressed.
Although energy is a more fundamental quantity than force, 
it is not directly accessible in the data analysis.
In the following, we first determine the force directly based on the time-series data and then introduce energy as the potential of the force.

The introduction of the force follows Newtonian mechanics.
In other words, we first define motion without force and then introduce motion caused by a force as a deviation from it. 
The force is then introduced through the physical laws that govern motion.
All of these procedures are carried out through quantities that can be measured by the Langevin equation, and we require general covariance in the expression of the laws to make them physically meaningful.

First, we identify the motion without force with the free Brownian motion.
In other words, we do not consider a random force as a "force" because they are not associated with the FEL. 
Unlike the motion of Brownian particles, the definition of Brownian motion for reaction coordinates requires invariance with respect to variable transformations in the definition of motion.

As explained, the Langevin equation cannot be expressed based on measured quantities; rather, the Fokker--Planck equation is more suitable for defining Brownian motion in the reaction coordinate.
Considering the invariance, the Fokker--Planck equation for Brownian motion is introduced as
\begin{align}
\pder{{\cal P}(z,t)}{t}
=\Delta {\cal P}(z,t).\label{DiffusionEquation}
\end{align}
Here, ${\cal P}(z,t)$ and $\Delta$ are defined as 
\begin{align}
{\cal P}(z,t)&\equiv P(z,t)\sqrt{\det {\bf D}(z)}, 
\label{invariant:P}\\
\Delta \phi(z)&\equiv \sqrt{\det {\bf D}(z)}\pder{}{z^i}
\left(\frac{D^{ij}(z)}{\sqrt{\det {\bf D}(z)}}
\pder{\phi(z)}{z^j}\right),
\label{LaplaceBeltrami}
\end{align}
where $\phi$ is an arbitrary scalar function.
$\cal P$ is scalar, i.e., invariant under the coordinate transformation.
$\Delta$ is the Laplace--Beltrami operator.
The invariance above allows us to obtain a physical definition of the force-free motion of the reaction coordinates.

Then, motion under force is introduced as a deviation from Eq. \eqref{DiffusionEquation}.
This deviation is expressed by the contravariant vector $u^i(z)$ in 
\begin{align}
\pder{{\cal P}(z,t)}{t}
=\Delta {\cal P}(z,t)
-\nabla_i (u^i(z) {\cal P}(z,t)).
\label{invariant:FokkerPlanck}
\end{align}
Here, the operator $\nabla_i$ is the covariant divergence defined as
\begin{align}
\nabla_i a^i(z) \equiv\sqrt{\det {\bf D}(z)}
\pder{}{z^i}(\frac{a^i(z)}{\sqrt{\det{\bf D}(z)}}),
\label{CovariantDivergence}
\end{align}
where $a^i$ is an arbitrary contravariant vector.
Eq. \eqref{FokkerPlanck} is rewritten as Eq. \eqref{invariant:FokkerPlanck} by expressing 
$u^i$ in terms of $v^i(z)$ and $D^{ij}(z)$ as
\begin{align}
u^i(z)=v^i(z) -\sqrt{\det{\bf D}(z)}\pder{}{z^k}(\frac{D^{ik}(z)}{
\sqrt{\det{\bf D}(z)}}).
\label{def:u}
\end{align}
Because $v^i$ and $D^{ij}$ can be evaluated from time-series data, we can determine whether $u^i$ is zero or not, that is, the presence or absence of force, from Eq. \eqref{def:u}.
Furthermore, the motion under the force is independent of the coordinate, i.e., physical, because $u^i(z)$ is a contravariant vector.

Now, the force can be determined as a covariant vector $f_i(z)$; otherwise, it cannot be expressed as a gradient of the potential.
Such a covariant vector can be introduced by a contravariant vector $u^i(z)$ and a metric tensor $g_{ij}(z)$ as follows
\begin{align}
f_j(z)\equiv g_{ij}(z)u^i(z),
\label{def:f}
\end{align}
where the metric tensor $g_{ij}$ is defined as 
\begin{align}
g_{ij}(z)D^{jk}(z)=\delta_i^k.\label{def:g}
\end{align}
If $z$ is a set of the thermodynamic extensive variables, Eq. \eqref{def:f} corresponds to the phenomenological deterministic dynamics in relaxation processes to the equilibrium state \cite{itami2017universal}.
Therefore, Eq. \eqref{def:f} should be taken as a well-known equation of motion rather than a newly introduced assumption.

Generally, the force, as a covariant vector, can be decomposed into three components as follows:
\begin{align}
f_jdz^j = d\psi+\delta a + b,
\label{hodge}
\end{align}
where $d$ and $\delta$ are the exterior derivative and the codifferential, respectively.
Eq. \eqref{hodge} is the Hodge decomposition for the 1-form, i.e., covariant vector.
Here, $\psi$, $a$, and $b$ are a scalar, 2-form, and harmonic form, respectively \cite{nakahara}.
The second and third terms represent the non-conservative contribution of the force.
The first term $d\psi$ is the potential contribution and $\psi$ will later become the FEL.
Since we are dealing with an equilibrium system, the second and third terms are required to be zero: $\delta a=b=0$.
Then, by integrating $f_j$, $\psi$ is expressed as a potential as follows:
\begin{align}
\psi(z)=\int
g_{ij}(z)\Big(v^i(z)
-\sqrt{\det {\bf D}(z)}\pder{ }{z^k}
(\frac{D^{ik}(z)}{\sqrt{\det {\bf D}(z)}})\Big)dz^j.
\label{psi:no:db}
\end{align}
According to this formula, we can see that $\psi$ is a functional of $v^i$ and $D^{ij}$.

Instead of assuming $\delta a=b=0$, 
the equilibrium condition can be expressed by the detailed balance condition, which is a consequence of the reversibility of microscopic motion \cite{van1992stochastic}.
Because we consider only even variables with respect to time reversal, the detailed balance condition is reduced to
\begin{align}
\left(v^i(z)-\pder{}{z^j}D^{ij}(z)\right)P(z)=0.
\label{detailedbalance}
\end{align}
It follows by straightforward calculation that the condition is invariant under the coordinate transformation, i.e., physical.
Substituting Eq. \eqref{def:u} into Eq. \eqref{def:f} with Eq. \eqref{detailedbalance},
we obtain the equation
\begin{align}
f_j(z)=&g_{ij}(z)\Bigg(\frac{1}{P(z)}\pder{}{z^k}(D^{ik}(z)P(z))\nonumber\\
&-\sqrt{\det {\bf D}(z)}\pder{}{z^k}\frac{D^{ik}(z)}{\sqrt{\det{\bf D}(z)}}\Bigg),\nonumber\\
=&\pder{}{z^j}\ln P(z)\sqrt{\det{\bf D}(z)}.
\end{align}
This equation appears to represent the existence of a potential 
\begin{align}
\psi(z)= \ln P(z)\sqrt{\det{\bf D}(z)}
+{\rm const.},
\label{def:psi}
\end{align}
Eq. \eqref{def:psi} is simpler than Eq. \eqref{psi:no:db} and convenient to evaluate from time-series data.
Here, $\psi$ is the Massieu potential whose arguments are generalized to the reaction coordinate.
Strictly speaking, the derivation of the potential $\psi$ is correct only if the manifold of reaction coordinates is contractible.
To show the existence of a potential on an arbitrary manifold requires a more sophisticated expression of the detailed balance condition, the so-called symmetrizable condition of the diffusion \cite{ikedawatanabe}.

\para{Requirement 3: Consistency}
Requiring consistency with statistical mechanics is essential for introducing the temperature into the expression of the potential $\psi$; otherwise, we cannot introduce an energy scale.

Before we look for consistency, a significant difference should be noted between the canonical distribution of $\Gamma$ and the probability distribution of $z$.
In the former case, both the Hamiltonian and the volume of the phase space are invariant with respect to the coordinate transformation. This is because coordinate transformation is a special case of canonical transformations.
This property makes the canonical distribution well defined.
In the latter case, on the other hand, there is no such invariance.
Despite its lack of invariance, the ill-considered use of the former procedure for the latter, written as Eq. \eqref{def1}, makes the conventional FEL unphysical.

When considering consistency with the canonical distribution, it is necessary to seek invariance of both the volume and probability distributions to coordinate transformations.
The invariance of volume elements comes from the way canonical momentum is constructed.
Because the canonical momentum is obtained from the generalized velocity and the mass-metric tensor, the origin of this invariance is the metric tensor. 
We, therefore, introduce an invariant volume element by the metric ${\bf g}(z)$ that appears in the Langevin dynamics of $z$ written as $dz\equiv \sqrt{\det {\bf g}(z)}dz_1dz_2...dz_n$.
Recall that $\bf g$ is the reciprocal of $\bf D$, and the normalization condition of the probability distribution can be written as
\begin{align}
\int dz {\cal P}(z)=1\label{normalize:iP}.
\end{align}
Here, we have used the scalar form of the stationary probability distribution 
${\cal P}(z)=P(z)\sqrt{\det{\bf D}(z)}$
introduced by Eq. \eqref{invariant:P}.

Now, we are ready to make a comparison with the canonical distribution.
We require that the ``physical'' FEL $F(z)$ be an effective Hamiltonian of the scalar probability distribution of $z$.
That is, we require that
\begin{align}
{\cal P}(z)=\frac{e^{-F(z)/T}}{Z}.
\label{def:FEL}
\end{align}
Here, $Z$ is a normalization constant of the probability distribution.
Then, Eq. \eqref{def:FEL} is rewritten as our main result Eq. \eqref{FEL}.
This is the formula of the physical FEL used to calculate from time-series data analysis.
If ${\bf D}(z)$ is independent of $z$, $F(z)$ in Eq. \eqref{FEL} is reduced to the conventional FEL $\Fc (z)$ in Eq. \eqref{def1}.

Eq. \eqref{FEL} contrasts the conventional FEL which does not restrict the choice of $z$ even though the effective description of $z$ is expected.
Eq. \eqref{FEL} is composed of three quantities: $T$, ${\bf D}(z)$, and $P(z)$.
These quantities correspond to thermodynamics, dynamics, and equilibrium condition, respectively.
The role of temperature is clear.
Without it, we cannot introduce quantities with dimensions of energy.
The roles of the latter two should be mentioned in detail.
In the over-damped Langevin description, dynamics is completely characterized by $v^i(z)$ and $D^{ij}(z)$, such that any physical quantity, including force, must be expressed by a combination of them.
Therefore, $P(z)$ in Eq. \eqref{FEL} is not necessarily required to introduce the FEL. Rather, $P(z)$ is a consequence of the detailed balance condition, which is a requirement of dynamics and eliminates $v^i(z)$ from $f_j(z)$ or $\psi$ in Eq. \eqref{psi:no:db}.
The expression of the FEL by dynamical characteristics must precede the simplified formula by equilibrium distributions, and they precede the consistency of the canonical distribution.
The failure of the conventional FEL was that the definition was given in reverse order.

We finally note that it is straightforward to extend our theory to systems described by the under-damped Langevin equation.
We only need to properly treat the degenerated diffusion matrix and variables that have odd parity with respect to time reversal. 
Such an extension is useful when treating, e.g., the Kramers' turnover.

\para{Discussion}
Fundamental physical objects, such as Lagrangian, Hamiltonian, and action, are always scalar.
Furthermore, physical law must be expressed in a covariant fashion, that is, in terms of tensor fields.
These rules must be followed in the expression of the FEL and the physical law that gives it; otherwise, the predictive ability as a physical law will be lost.
Unfortunately, the conventional FEL does not adhere to this principle.
We have achieved this requirement by expressing it as in Eqs. \eqref{def:f}, \eqref{detailedbalance}, and \eqref{def:FEL}, and we then obtained
 the FEL in Eq. \eqref{FEL} that has the same predictive ability as other physical laws.

Eq. \eqref{FEL} is not a proposal for yet another definition of the FEL, but a formula derived from physical requirements.
It properly contains the conventional FEL as a special case.
Thus, there is no room for arbitrariness and no other choice.

We have so far assumed that a set of reaction coordinates that is sufficient to describe the reaction of interest is already known.
For the set of coordinates, the Arrhenius rate should show good agreement with a directly measured reaction rate.
In other words, this agreement provides a criterion to evaluate how properly the set of reaction coordinates describes a reaction path.
According to the previous definition, disagreement indicates two possibilities: the variables necessary to describe the reaction were not taken into consideration or the metric was not properly selected.
Our formulation eliminates the latter possibility, such that when redefining an alternative set of reaction coordinates we can focus only on improving the problems arising from the former possibility.
Data-driven methods will help to solve the former problem efficiently; therefore, they are complementary to our formulation
\cite{abhinav2010hidden,inoue2006analysis,takano1995relaxation,kawai2015environmental}.

We expect that our formula for the FEL will enable us to develop a material design method aided by simulations that is considerably more effective than those based on the conventional unphysical FEL.
The FEL may also be applied to detect energy transfer, e.g., in molecular motor experiments, which is hardly analyzed within the conventional schemes.

\begin{acknowledgments}
The author thanks Shin-ichi Sasa, Yohei Nakayama, Masato Itami, Yuki Minami, Macoto Kikuchi, Tetsuya Morishita, Masakazu Matsumoto, Hiroshi Fujisaki, Motoyuki Shiga, and Wataru Shinoda for stimulating discussions and Hal Tasaki for a critical reading of the manuscript.
This work was supported by JSPS KAKENHI Grant Number 15K13530, 19H01864, and 18H01188 and JST-PRESTO Grant Number 
JPMJPR15ND, Japan.
\end{acknowledgments}

\end{document}